\documentclass[reprint,aps,prl,superscriptaddress,amsmath,amssymb,floatfix,footinbib,longbibliography]{revtex4-1}
\usepackage{graphicx}
\usepackage{dcolumn}% Align table columns on decimal point
\usepackage{bm}% bold math
\usepackage{amsmath}
\usepackage{times}
\usepackage{color}
\usepackage[breaklinks=true,colorlinks,citecolor=blue,linkcolor=blue,urlcolor=blue]{hyperref}

\makeatletter

\newcommand{\Rmnum}[1]{\expandafter\@slowromancap\romannumeral #1@}
\makeatother

\begin{document}

\title{Junction-free polarity-tunable superconducting diode effect enables XNOR logic operation}

\author{Alapan Bera}
\affiliation{Department of Physics, Indian Institute of Technology Kanpur, Kanpur 208016, India}
\author{Soumik Mukhopadhyay}
\email{soumikm@iitk.ac.in}
\affiliation{Department of Physics, Indian Institute of Technology Kanpur, Kanpur 208016, India}

\begin{abstract}
The realization of the superconducting diode effect (SDE), characterized by non-reciprocal supercurrent, is a major step towards dissipationless logic circuits. A polarity-tunable superconducting diode with high diode efficiency primarily consists of Josephson-junction architectures that couple two superconducting regions via a weak link. Realizing an SDE without a weak-link junction would reduce fabrication complexity, lower electrical noise, and minimize dependence on junction quality. Here, we report an all-vdW, highly efficient, junction-free SDE in NbSe$_2$, enabled by proximity-induced symmetry breaking from a strong uniaxial ferromagnet Fe$_3$GeTe$_2$ (F3GT). Superconducting diode efficiency of up to $30\%$ is achieved at a small magnetic field of 50 mT. A field-free diode rectification effect is realized with diode polarity that is tunable by the F3GT spin polarization. Furthermore, by employing the active-field dependence of the SDE, a two-input exclusive-NOR (XNOR) logic gate operation is demonstrated. The findings highlight the potential utilization of magnetic proximity effect towards realizing junction-free superconducting diodes for high-performance, energy-efficient logic circuits.   

\end{abstract}

\maketitle

{\it Introduction.---}
Non-reciprocal ultra-low resistive circuit elements hold enormous promise for the dissipation-less, low-power computing and information processing at cryogenic temperatures. The superconducting diode effect (SDE), which allows a unidirectional flow of a zero-resistance superconducting current~\cite{Nadeem2023, dibernardo2026, Sarkar2026}, has been a proposed route for such purposes. The non-reciprocity of supercurrent in SDE is fundamentally driven by broken inversion and/or time-reversal symmetry, which may be achieved in non-centrosymmetric quantum materials and in the presence of magnetic interactions~\cite{Nadeem2023}. Despite longstanding theoretical predictions, SDE has been experimentally realized only recently, predominantly in artificially engineered non-centrosymmetric multilayers involving superconductors and Josephson junction architectures~\cite{Ando2020, Jeon2022,adfm.202311229,Pal2022,Wu2022,doi:10.1126/science.abl8371,PhysRevLett.130.266003,Diez-Merida2023,Ghosh2024,Wu2025,doi:10.1126/sciadv.adw6925,Ingla-Aynes2025,https://doi.org/10.1002/adma.202513434} comprising a wide range of material systems. While a Josephson diode can leverage an artificially broken inversion symmetry at the junction interfaces and time-reversal symmetry (TRS) breaking with a magnetic barrier, it does have several technical drawbacks due to factors like complex weak-link fabrication, sensitive dependency on weak-link characteristics, parasitic capacitance, junction noise. etc. These effects can hinder the diode performance significantly, leading to the necessity of alternative device architectures.

A junction-free superconducting diode, where the non-reciprocity is realized in a monolithic superconducting layer, is a promising candidate to bypass these geometric complexities. The required symmetry breaking can achieved by tuning the intrinsic material characteristics and/or proximity-driven phenomena induced by adjacent layers. Most of the studies reporting junction-free SDE incorporate complex device geometries~\cite{Bauriedl2022,Lyu2021}, precisely-controlled twist angles~\cite{Lin2022} or multilayer stacked architectures~\cite{Narita2022, https://doi.org/10.1002/adfm.75836,https://doi.org/10.1002/adma.202511414}, while some others report an intrinsic SDE with low diode efficiency~\cite{Han2026,Le2024,Qi2025} or degradation issues~\cite{PhysRevB.90.081402}. Van der Waals (vdW) materials, with their atomically smooth surface and superior interface quality in heterostructure devices, can play a pivotal role in this direction to enable simpler device structures with high diode efficiency. Discoveries of Nb- and Fe-based vdW superconductors~\cite{Hamill2021, Wang2017, PhysRevResearch.4.013188, Cho2021, Ge2015, PhysRevLett.130.046702, sym12091402, Qiu2023} and Fe-based novel vdW ferromagnets~\cite{Fei2018, Tan2018, 10.1021/acsami.6c08036, Zhang2022,doi:10.1126/sciadv.aay8912, wmb3-5r6b, https://doi.org/10.1002/pssr.70156, 10.1021/acsnano.2c01948, PhysRevB.110.224401, PhysRevMaterials.3.104401} further elevate the potential. Recent realizations of Josephson junctions (JJ) and superconducting quantum interference devices (SQUID) based on vdW superconductors like NbSe$_2$ ~\cite{Ma2025, Wu2022, Wu2025}, NbS$_2$~\cite{PhysRevResearch.6.L012046}, double-layer graphene~\cite{doi:10.1021/acs.nanolett.0c02412}, and vdW magnetic layers like Cr$_2$Ge$_2$Te$_6$~\cite{doi:10.1021/acs.nanolett.2c01640, Ai2021}, MnBi$_2$Te$_4$~\cite{Jansen2024, doi:10.1126/sciadv.ads8730}, NiPS$_3$~\cite{gonzalezsanchez2025}, Fe$_3$GeTe$_2$~\cite{Hu2023, https://doi.org/10.1002/adma.202513434}, underscore the significance of vdW-based superconducting devices to give rise to technologically significant results. However, to date, the prospect of such junction-free, highly efficient vdW-based SDE remains largely unexplored. Very recently, a few works on vdW superconductor/magnet heterostructures~\cite{https://doi.org/10.1002/adfm.202504056,PhysRevResearch.5.L022064,doi:10.1021/acsami.5c19869, Hu2025} demonstrated a magnetic proximity-induced SDE. However, they still lack a diode efficiency comparable to that of the Josephson junctions and/or fail to achieve a polarity-reversible zero-field diode effect due to anti-ferromagnetic proximity layers. Furthermore, the implementation of a fully functional logic gate operation, based on such a junction-free SDE, still remains to be illustrated.  

Here, we report a highly-efficient SDE in the Ising superconductor NbSe$_2$, originating from the magnetic proximity of a strong uniaxial ferromagnet Fe$_3$GeTe$_2$ (F3GT). We present an all-vdW Fe$_3$GeTe$_2$/NbSe$_2$/Fe$_3$GeTe$_2$ (F/S/F) heterostructure to reveal a polarity-tunable field-free diode effect, induced by remnant magnetization in F3GT. A large diode efficiency of nearly 30\% is obtained at a small field of 50 mT, comparable to that of the Josephson junction counterparts. The zero-field diode rectification effect is presented with controllable diode characteristics using the spin configuration in F3GT. Additionally, we demonstrate an exclusive-NOR (XNOR) logic gate operation that incorporates the field-tunable diode polarity, enabling energy-efficient logic devices for superconducting electronics. 

\begin{figure*}[htp]
\includegraphics[width=0.8\linewidth]{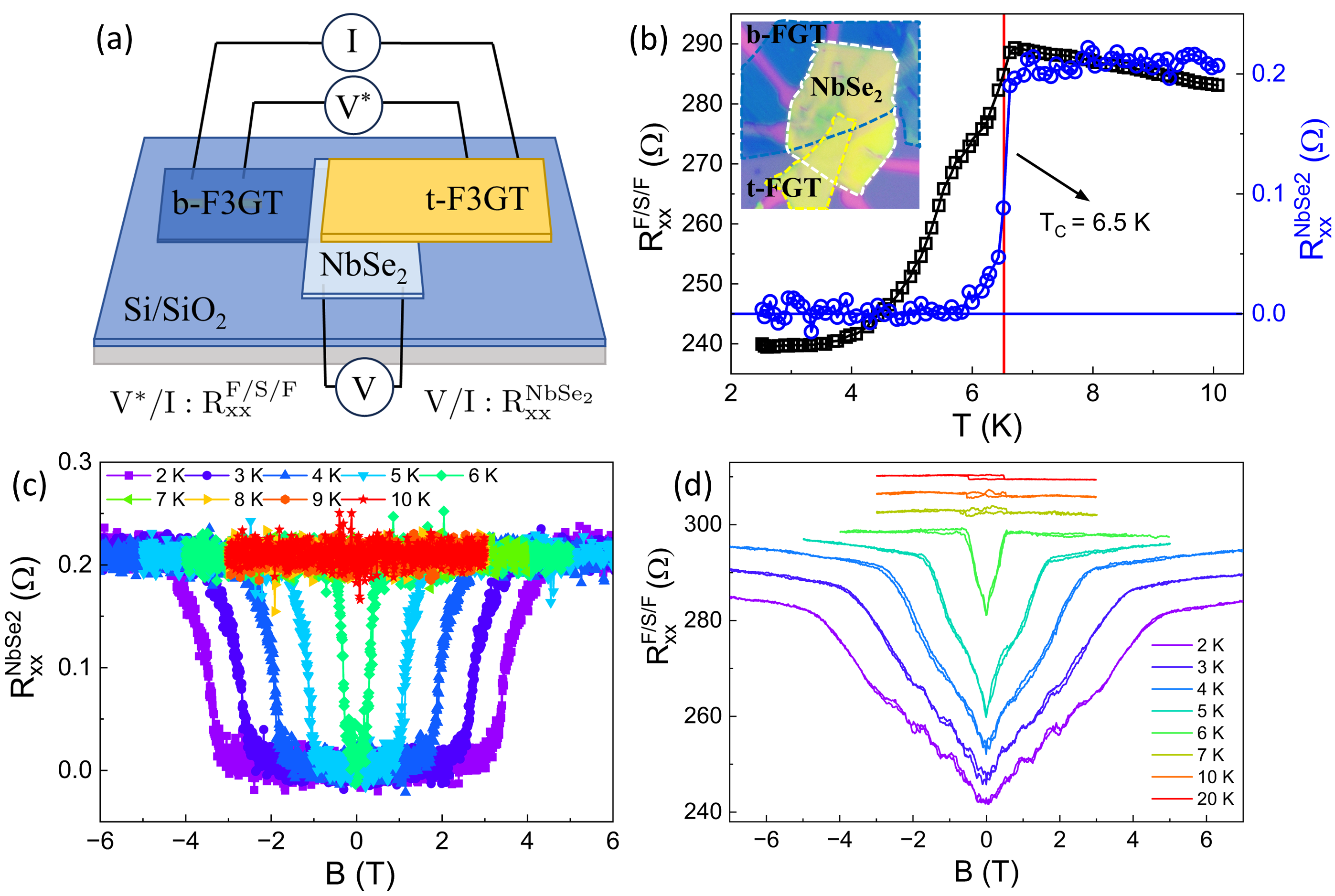}
\caption{(a) Schematic diagram of the F/S/F device architecture, along with the electrical transport measurements protocol. The junction-free SDE is studied in the bare NbSe$_2$ layer, as denoted by the voltage contact probes in V. (b) Temperature dependence of the electrical resistance across the F/S/F junction (black symbol), as well as in the NbSe$_2$ layer (blue symbol). An optical image of the measured device is shown in the inset. (c) Magnetic-field dependence of the NbSe$_2$ resistance at various temperatures, measured for B$\parallel$c configuration. The critical field, marking a transition from the superconducting state to a normal state, decreases to zero as the critical temperature is approached. (d) The temperature-driven evolution of MR curves across the F/S/F junction. A multi-fold increase in the MR ratio is observed as the measurement temperature drops below 6.5 K.} 
\label{fig1}
\end{figure*}  

{\it Results and discussion.---} Single crystal of the ferromagnetic F3GT and superconductor NbSe$_2$ are grown using the chemical vapor transport method. The nanoflakes of these materials can be exfoliated mechanically, as both systems exhibit a vdW crystal structure with layer-by-layer growth. See the supporting information~\cite{SM} for details of sample growth and characterization. NbSe$_2$ is a type-II superconductor, with a superconducting critical temperature around $7~\mathrm{K}$~\cite{PhysRevLett.98.057003, Xi2016}. It is known for exhibiting an Ising superconductivity~\cite{Xi2016} in the low-thickness limit, which arises from a combined effect of non-centrosymmetric crystal structure and spin-orbit interaction. F3GT, on the other hand, is a hard ferromagnet known for a strong perpendicular magnetic anisotropy (PMA) and an elevated Curie temperature~\cite{10.1021/acsami.6c08036,Fei2018} among vdW ferromagnets. From the DC magnetization measurements, we obtain a Curie point of $\mathrm{T_C^F}= 233~\mathrm{K}$ and a PMA constant of $\mathrm{K_U}=1.56\times10^6~\mathrm{J/m^3}$ (refer to Sec. II of the supporting information~\cite{SM} for details of magnetization measurement).

\begin{figure*}[htp]
\includegraphics[width=0.85\linewidth]{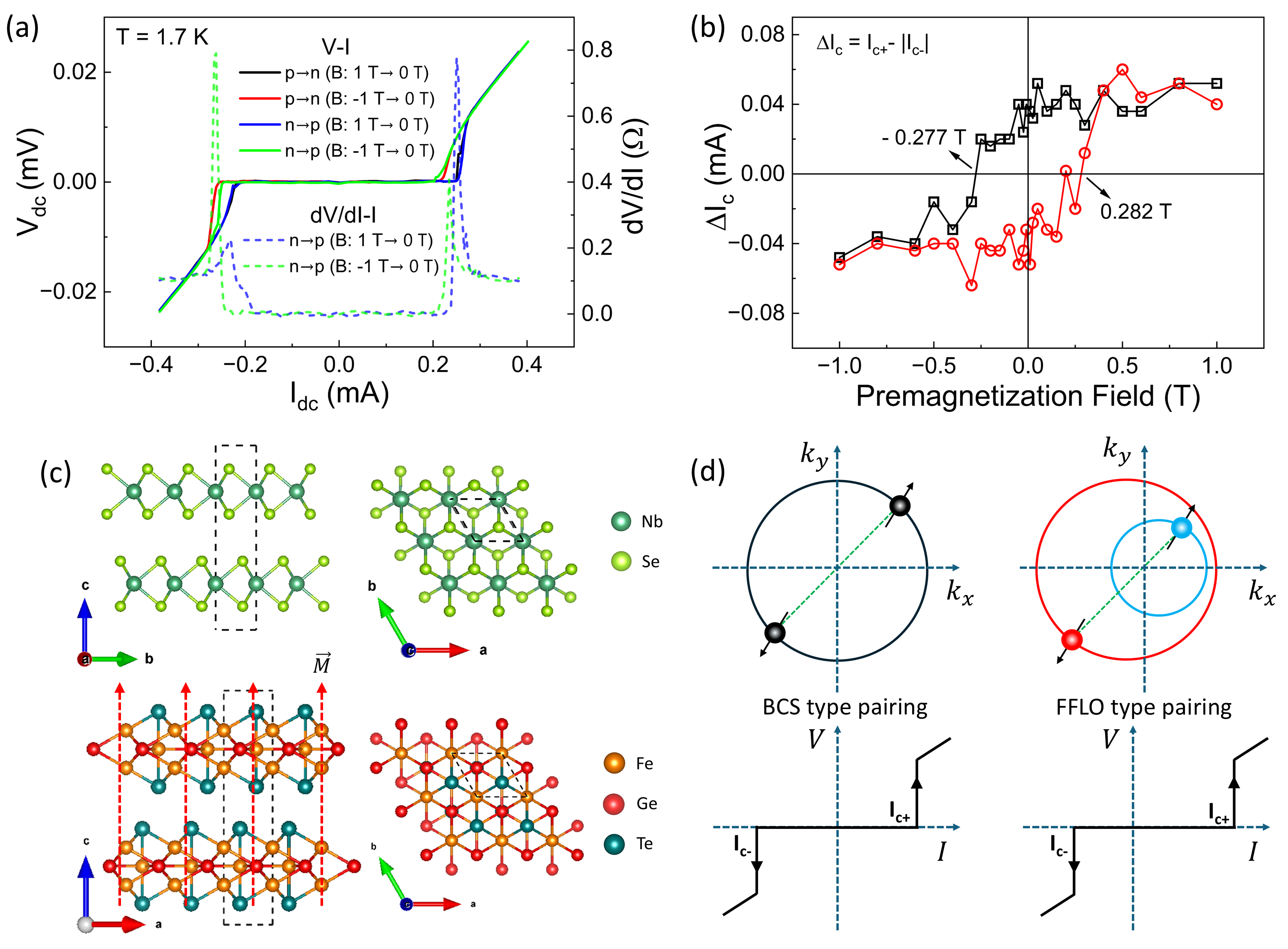}
\caption{(a) The V-I and dV/dI-I characteristics, measured in the NbSe$_2$ layer under different premagnetization fields, are represented by the solid and dashed curves, respectively. An asymmetric current dependence is visible in all cases, revealing a 
proximity-induced non-reciprocal transport.  The reversal from a higher positive critical current ($\mathrm{I_{c+}}$) at 1 T premag field to a higher negative critical current ($\mathrm{I_{c-}}$) at -1 T premag field is evident, due to a flipped spin configuration at F3GT. (b) The non-reciprocal part of the critical currents ($\Delta \mathrm{I_c}$), plotted as a function of various premag fields, shows a hysteresis pattern with a coercivity of 0.28 T. The black and red symbols display positive-to-negative and negative-to-positive field-setting sequences, respectively. (c) Side view (left panel) and top view (right panel) of the crystal structure of NbSe$_2$ (top panel) and F3GT (bottom panel), illustrating the hexagonal lattice symmetry associated with the P6$_3$/mmc space group. The remnant magnetization orientation along the c-axis in the F3GT layer is denoted by red arrows. (d) Top panel: schematic diagram of the Cooper pair formation in BCS-type pairing and FFLO-type pairing. A non-zero center-of-mass of the Cooper pairs emerges in the FFLO picture in the presence of inversion and time-reversal symmetry breaking. Bottom panel: the finite momentum Cooper pairs lead to a non-reciprocity in the critical current (right) in the FFLO-type pairing, in contrast to a BCS superconductor (left).}
\label{fig2}
\end{figure*} 

\begin{figure*}[htp]
\includegraphics[width=0.85\linewidth]{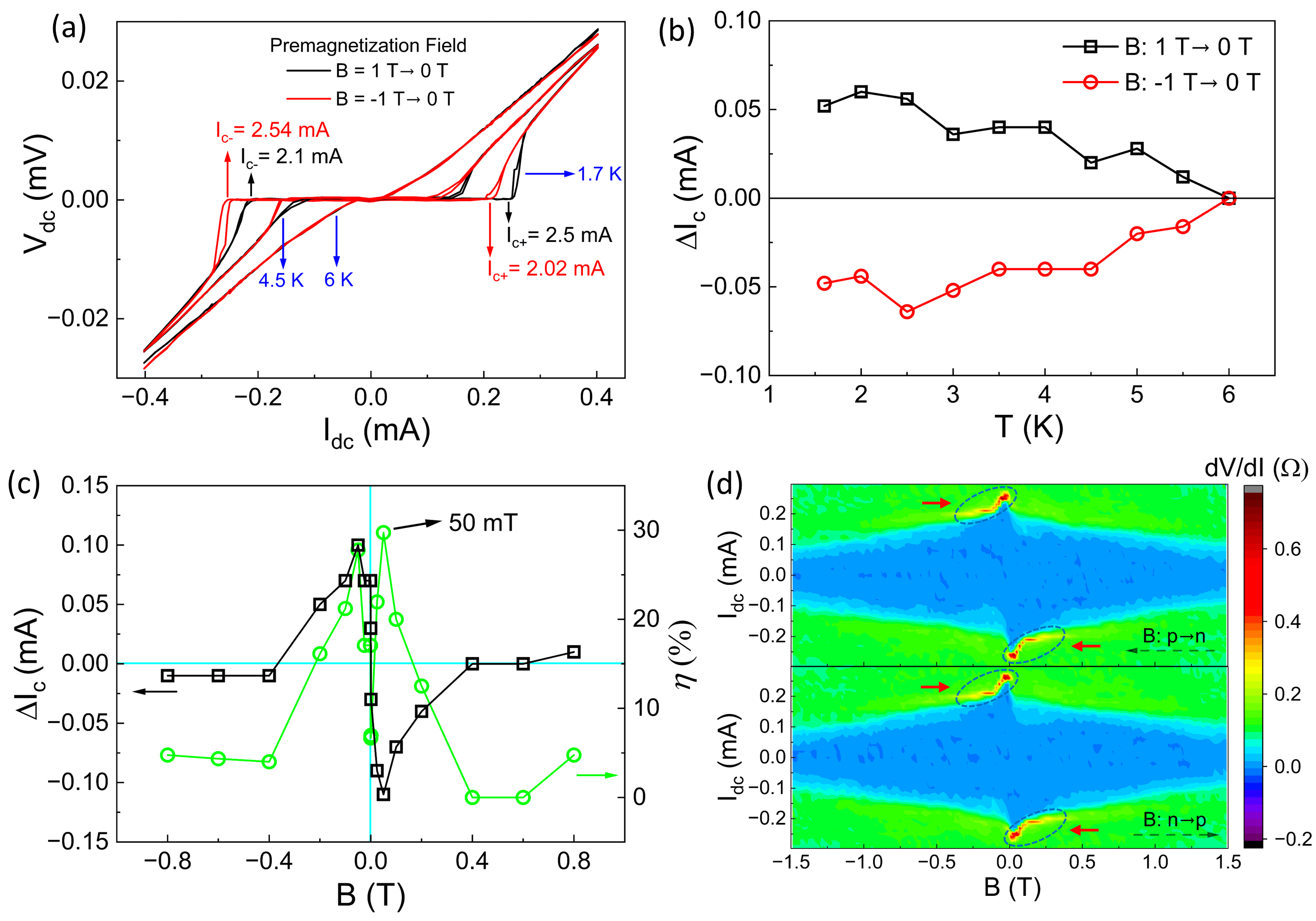}
\caption{(a) Temperature-driven modulation of the diode behavior, measured at premagnetization fields of 1 T and -1 T, is shown in the black and red curves, respectively. (b) Temperature dependence of the $\Delta \mathrm{I_c}$ for positive and negative premag field of 1 T, highlighting that the robust SDE is persistent up to 5.5 K. (c) Active magnetic field dependency of $\Delta \mathrm{I_c}$ and diode efficiency $\eta$, as calculated using Eq.~\ref{Eq1}. A maximum efficiency of 29.7\% is observed at $\pm$50 mT. (d) Contour plot of the active magnetic field and current-bias dependency of differential resistance. A clear asymmetric response to the current flow direction is observed, with minimal sensitivity to the field-sweep protocol.}
\label{fig3}
\end{figure*} 

To study the proximity effect of the strong PMA ferromagnet F3GT on the superconducting phase of NbSe$_2$, we fabricate a trilayer F/S/F device structure. A NbSe$_2$ nanoflake (35 nm) is sandwiched between two F3GT layers of thickness 18 nm and 231 nm. Refer to the supporting information~\cite{SM} for the details of thickness determination of the nanoflakes using atomic force microscopy (AFM). F3GT flakes with significantly different thicknesses are selected to introduce structural asymmetry across the NbSe$_2$ layer. Metallic contact probes are drawn from all three regions, as shown in the device schematic in Fig.~\ref{fig1}(a), to probe the electrical transport across the F/S/F junction (V$^*$) and the bare NbSe$_2$ layer (V) simultaneously. Inset of Fig.~\ref{fig1}(b) presents an optical image of the device under test. Temperature dependencies of the longitudinal electrical resistance ($R-T$) across the F/S/F junction (black symbols) and the NbSe$_2$ layer (blue symbols) are shown in Fig.~\ref{fig1}(b). The NbSe$_2$ layer exhibits a sharp normal metal-superconductor transition at $\mathrm{T_C^S}=6.5$ K, which is marginally lower compared to the bulk counterpart~\cite{PhysRevLett.98.057003, Xi2016}. On the other hand, the $R-T$ curve across the F/S/F junction shows a gradual reduction in resistance that starts around the same temperature and saturates below 4 K. This resembles a Berezinskii-Kosterlitz–Thouless (BKT) transition~\cite{Hu2023} that stems from proximity-induced vortex-antivortex pairs at the junction. Note that the junction resistance settles to a nonzero value at the lowest temperature due to finite contributions from the F3GT layers. 

The temperature dependence of the magneto-resistance (MR) curves, measured in the bare NbSe$_2$ and across the F/S/F junction, is presented in Fig.~\ref{fig1}(c) and (d), respectively. The magnetic field (B) is applied along the c-axis, parallel to the easy-axis of F3GT. A typical superconductor-normal metal phase transition is observed in the NbSe$_2$ layer [Fig.~\ref{fig1}(c)], where the lower critical field $\mathrm{H_{C1}}$ gradually reduces to zero near $\mathrm{T_C^S}$. The junction magneto-resistance, on the other hand, exhibits a continuous rise with field up to the upper critical field $\mathrm{H_{C2}}$, at all temperatures below $\mathrm{T_C^S}$. This behavior resembles a proximity-driven multi-fold MR enhancement reported earlier~\cite{doi:10.1021/acsami.3c15363}, which is described by spin-triplet superconductivity with a split spin-band. In our study, however, a continuous increment in the MR ratio is observed with decreasing temperature, which is in agreement with the vortex-creation picture reflected in $R-T$. This bypasses the tunneling mechanism, leading to the absence of any MR plateau. A hysteresis pattern emerges in the MR curve above the critical temperature, where the MR ratio becomes almost negligible. This arises from a finite Hall response contribution of F3GT, due to a marginal probe misalignment error. The hard ferromagnetic nature is reflected in the hysteresis, which gets suppressed at low temperatures due to proximity-induced superconductivity in F3GT~\cite{doi:10.1021/acsnano.4c16050}. 

\begin{figure*}[htp]
\includegraphics[width=0.85\linewidth]{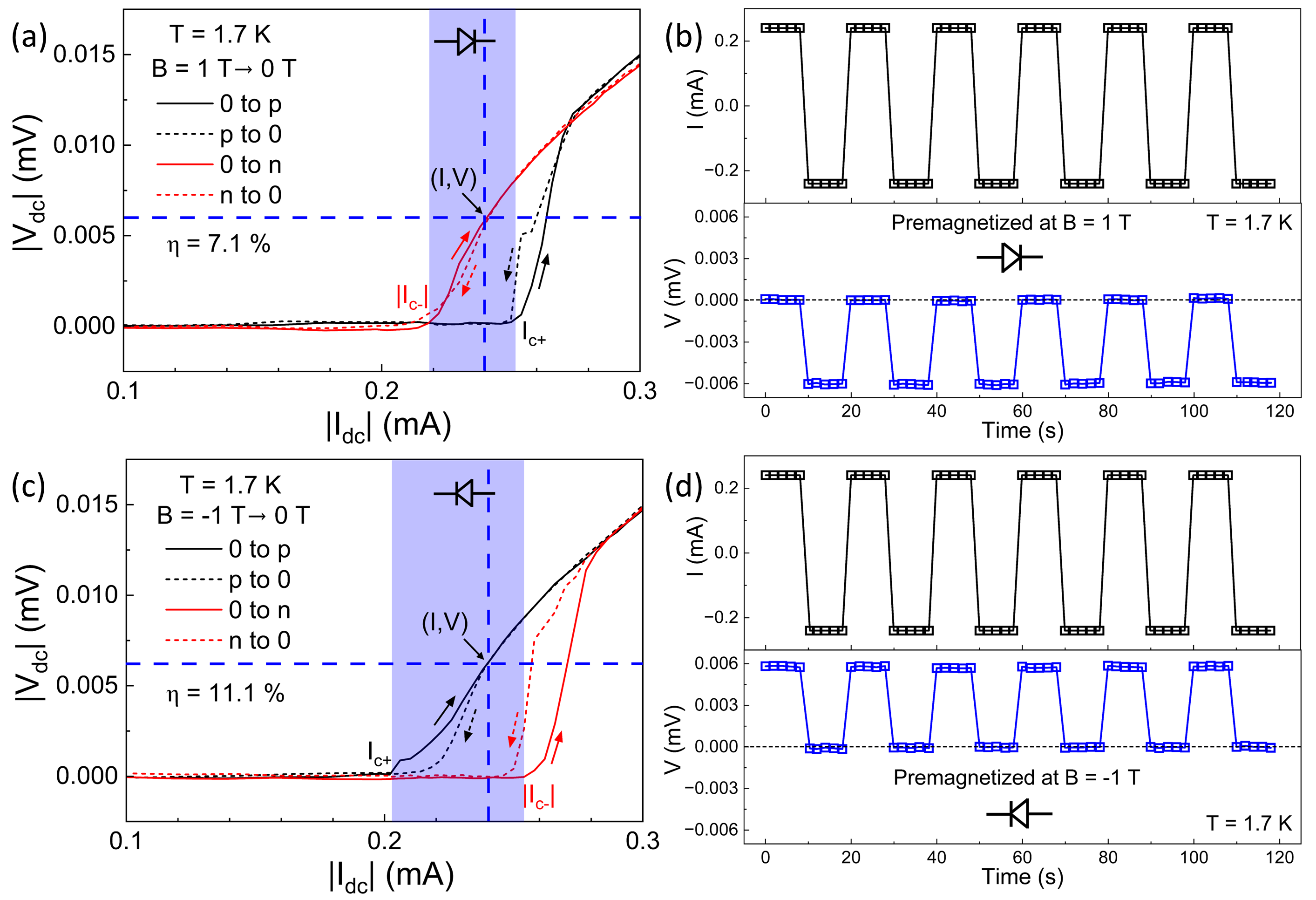}
\caption{(a) The V-I characteristics at a premag field of 1 T, where the negative-current quadrant is superimposed on the positive half with a modulus operation. The non-reciprocal region is shaded in purple. (b) The diode rectification effect is demonstrated using a square-wave input with a 0.24 mA amplitude and 0.05 Hz frequency. A zero-voltage output is obtained during the positive cycle, indicating dissipationless transport. (c) The V-I characteristics at -1 T premag field, with a reversed SDE polarity. (d) A corresponding reversed rectification process is observed with a supercurrent flow for negative cycles. The dashed lines in (a) and (c) indicate the operating current and the resulting voltage levels for resistive transports.} 
\label{fig4}
\end{figure*} 

\begin{figure*}[htp]
\includegraphics[width=\linewidth]{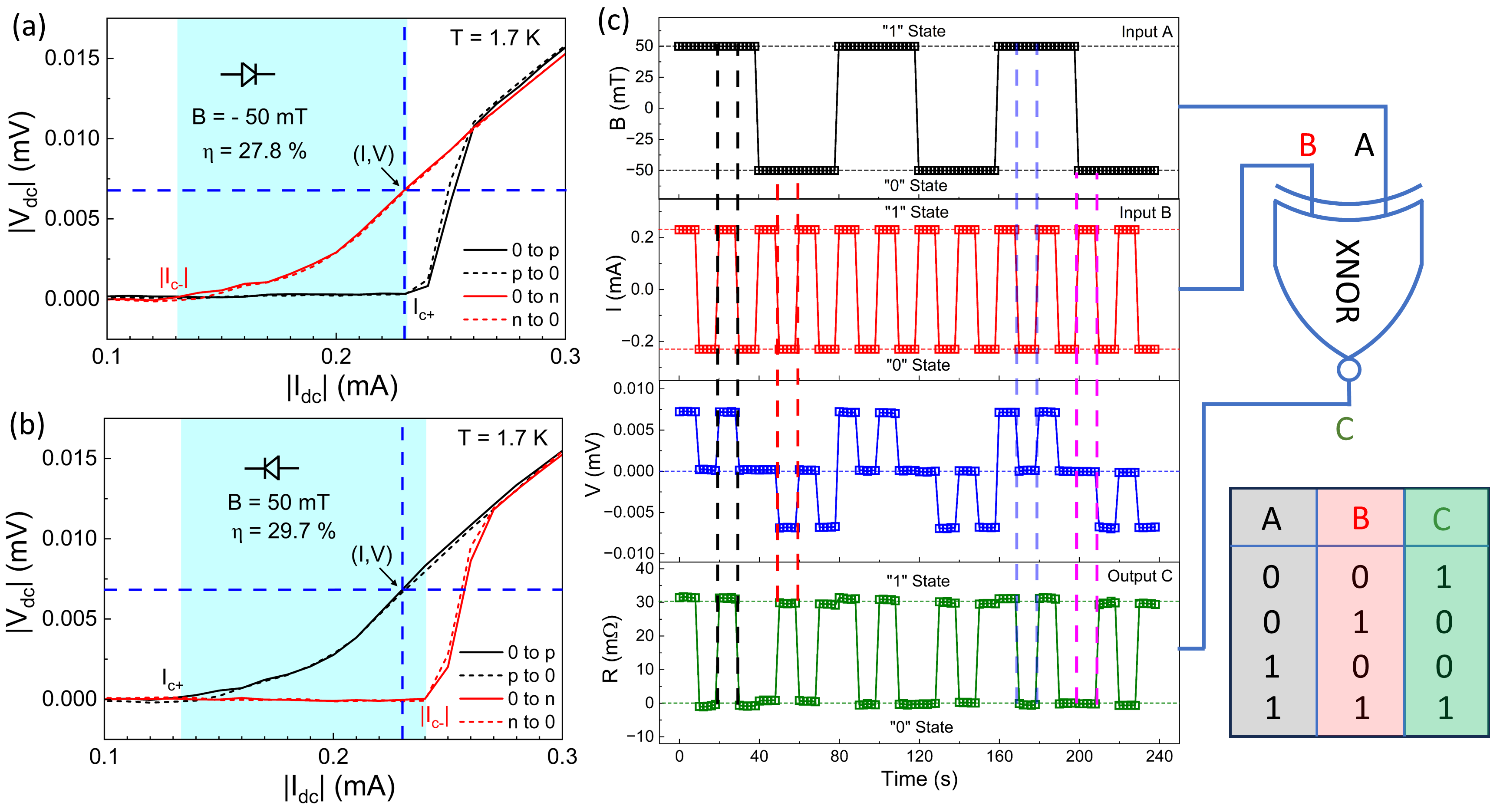}
\caption{The V-I characteristic in the presence of (a) -50 mT and (b) 50 mT magnetic fields. A large enhancement in the diode efficiency is observed, close to the Josephson junction benchmarks. The non-reciprocal region is highlighted in cyan, lying across a broad range of 100 $\mu$A. The intersection of the dashed lines represents the operating point of the following XNOR logic measurements. (c) The active field value is used as input A, and a square wave is used as input B, to demonstrate an XNOR logic operation with the resultant resistance as the output C. The first panel (from top) illustrates the applied time-varying magnetic field with an amplitude of 50 mT and a time period of 40 seconds. The second panel plots the 230 $\mu$A square-wave input current with a time period of 20 seconds. The resulting voltage and resistance outputs are shown in the bottom panels, which are consistent with the live-field SDE. The dashed lines of various colors correspond to the 4 distinct input/output combinations, as shown in the truth table at the bottom right.} 
\label{fig5}
\end{figure*} 

From here onward, we present a detailed study of the junction-free magneto-electrical transport properties in the NbSe$_2$ layer. Current-bias dependency of the voltage (V-I) and differential resistance (dV/dI-I) curves are shown in Fig.~\ref{fig2}(a), plotted in the solid and dashed lines, respectively. These characteristics are measured at a temperature of 1.7 K, under various magnetic-field protocols to manipulate the spin configuration of the adjacent F3GT layers. The B: 1 T$\rightarrow$ 0 T and B: -1 T$\rightarrow$ 0 T conditions in Fig.~\ref{fig2}(a) denote that the measurements are recorded for premagnetization (premag) fields of 1 T and -1 T, respectively. The premagnetization process involves applying a magnetic field of 1 T in parallel to the uniaxial anisotropy axis (c-axis), followed by the field withdrawal. The large field sets an upward spin alignment in F3GT, resulting in a remnant spin-up state at a zero active field. Similarly, a zero-field spin-down state is induced using a -1 T premag field sequence. Fig.~\ref{fig2}(a) presents the V-I curves for these premag conditions, as well as for opposite current sweep protocols. Transition from a dissipationless superconducting state to a resistive normal metal state is observed at some critical current in each scenario. We acquire the V-I curves with a 4-quadrant DC sweep with maximum current ($\mathrm{I_{max}}$) of 400 $\mu$A: $\mathrm{+I_{max}(p)} \rightarrow\mathrm{-I_{max}(n)}\rightarrow\mathrm{+I_{max}(p)}$. The critical currents for upward current sweep (0 to $\mathrm{\pm I_{max}}$) is defined as $\mathrm{I_{c\pm}}$ and for downward current sweep ($\mathrm{\pm I_{max}}$ to 0) is defined as $\mathrm{I_{r\pm}}$. Interestingly, for a premag field of 1 T, a significantly larger critical current is observed for the positive current direction, as compared to the negative half ($\mathrm{I_{c+}}>\mathrm{I_{c-}}$). This implies that an electrical current with an amplitude between $\mathrm{I_{c+}}$ and $\mathrm{|I_{c-}|}$ would face no resistance for a positive directional flow, while facing significant resistance for a negative flow direction. This is called the superconducting diode effect (SDE), which allows an unidirectional, dissipationless transport. The non-reciprocal component of the critical currents, defined as $\mathrm{\Delta I_c=I_{c+}-|I_{c-}|}$, turns out to be positive in this scenario. On the other hand, an opposite behavior is observed for a premag field of -1 T, with a negative $\mathrm{\Delta I_c}$ value. Notably, no significant difference between $\mathrm{I_{c}}$ and $\mathrm{I_{r}}$ is observed at any of the premag fields, suggesting an overdamped non-reciprocal transport~\cite{Ma2025, Ai2021}. The superconductor-normal metal transition is manifested in the pronounced peaks in dV/dI-I curves near $\mathrm{I_c}$. The SDE behavior is reflected through the contrast in the peak positions and peak amplitudes under reversed premag conditions.

The non-reciprocal component of the critical current $\Delta \mathrm{I_c}$ as a function of various premag fields is depicted in Fig.~\ref{fig2}(b). We set the premag conditions in two opposite sequences: (1) in a decreasing order from 1 T to -1 T (black symbols), and (2) in an increasing order from -1 T to 1 T (red symbols). Transition from positive $\Delta \mathrm{I_c}$ at 1 T to a negative $\Delta \mathrm{I_c}$ at -1 T is visible in both sweeps, with a saturating behavior at higher fields ($> 0.5$ T). The switching field ($\mathrm{B_{sw}}$), where a sign change in $\Delta \mathrm{I_c}$ takes place, is dependent on the premag sequence, leading to a hysteretic behavior that is consistent with the Hall response of F3GT [Fig.~\ref{fig1}(d)]. This is a direct verification of the diode effect originating from the magnetic-proximity effect induced by F3GT. We obtain a $\mathrm{|B_{sw}|}\approx0.28$ T, which is significantly larger than that reported in a Josephson junction~\cite{https://doi.org/10.1002/adma.202513434}, suggesting the robustness of the SDE. 

The superconducting diode effect is a symmetry-driven transport phenomenon and requires inversion and/or time-reversal symmetry breaking~\cite{Nadeem2023} for nonreciprocal depairing current. The device architecture inherently breaks down the inversion symmetry. Crystal structures of the NbSe$_2$ and F3GT systems are shown in Fig.~\ref{fig2}(c). The hexagonal lattice symmetry associated with the P6$_3$/mmc space group (for both systems) is evident from the top view (right panel). The NbSe$_2$/F3GT interface itself introduces the noncentrosymmetry required for the diode effect. Red (dashed) arrows in Fig.~\ref{fig2}(c) indicate the remnant magnetization of the F3GT layer along the crystallographic c-axis. Magnetic proximity of the strong PMA ferromagnet induces time-reversal asymmetry in NbSe$_2$, resulting in spin-split bands. The combined effect of inversion and time-reversal symmetry breaking results in a finite momentum Cooper pair state, in contrast to the conventional Bardeen-Cooper-Schrieffer (BCS) picture. Fulde–Ferrell–Larkin–Ovchinnikov (FFLO)-type Cooper pairing is well-known for such symmetry conditions, where the Zeeman interaction leads to spin splitting at $\mathbf{K}$ and $\mathbf{K^{\prime}}$ points and results in a finite center-of-mass momentum spin-singlet Cooper pair. The pairing mechanism in the BCS and FFLO-type superconducting state is illustrated in Fig.~\ref{fig2}(d). Imbalance in positive and negative momentum Cooper pairs leads to the non-reciprocal critical current associated with SDE (lower panel). Additionally, recent studies of F3GT and NbSe$_2$ based heterostructures point towards the possibility of ferromagnetic proximity-driven spin-triplet Cooper pairing formation, which has led to observations like improved diode efficiency~\cite{https://doi.org/10.1002/adma.202513434} and large MR enhancement~\cite{doi:10.1021/acsami.3c15363}. 

Fig.~\ref{fig3}(a) presents the evolution of the SDE with increasing temperature, with the 4-quadrant V-I curves plotted for both premag conditions. The diode behavior gets weaker at higher temperatures and disappears at 6 K. This temperature dependence is summarized in Fig.~\ref{fig3}(b), where an almost linear decrease in $\Delta \mathrm{I_c}$ is observed as the superconducting critical temperature is approached. This is a notable enhancement in the SDE resilience against temperature, compared to the Josephson junction counterpart that exhibits a finite $\Delta \mathrm{I_c}$ up to 4 K~\cite{https://doi.org/10.1002/adma.202513434}.
We further investigate the non-reciprocal behavior in the presence of an active (live) magnetic field along the c-axis. The active field-dependency of the non-reciprocal critical current $\mathrm{\Delta I_c}$ is presented in Fig.~\ref{fig3}(c), showing an antisymmetric and non-monotonic behavior in contrast to the premagnetization scenario. A strong non-reciprocity is observed at 0.5 m T magnetic field, with a $\Delta \mathrm{I_c}$ larger than 100 $\mu$A. The superconducting diode efficiency, a quantity for measuring the strength of the SDE, is given by: 
\begin{equation}
\mathrm{\eta}=\mathrm{\frac{|(I_{c+}-|I_{c-}|)|}{I_{c+}+|I_{c-}|}\times 100\%}.\label{Eq1}
\end{equation}
A high diode efficiency of 29.7 \% is observed at 50 mT, comparable to that of the Josephson junction diodes~\cite{https://doi.org/10.1002/adma.202513434}. Both the premag field-dependency and the active field-dependency of the SDE are measured at 1.7 K. Fig.~\ref{fig3}(d) presents the contour plot of the differential resistance as a function of the active magnetic field. The red patches depict large and sharp dV/dI peaks, associated with an abrupt transition to a resistive state at a higher supercurrent. The asymmetry in these red spots is evident in the contour plot, highlighting the non-reciprocity. No notable distinction is observed between the positive-to-negative (p$\rightarrow$n) and negative-to-positive (n$\rightarrow$p) field sweeps, as depicted in the upper and lower panels of Fig.~\ref{fig3}(d), underlining the absence of any hysteresis in the presence of an active field. 

We employ the premag field-controlled superconducting diode effect to demonstrate a zero-field diode rectification. Fig.~\ref{fig4}(a) presents the V-I curve at a 1 T premag field, with overlapped quadrants due to absolute current-voltage axes. The black and red curves present the negative and positive current quadrants, respectively. The non-reciprocal current regime is highlighted in purple, leading to a diode efficiency $\eta$ of 7.1\%. An associated diode rectification effect is depicted in Fig.~\ref{fig4}(b). An input current of square waveform is applied, with 0.24 mA amplitude and a time period T = 20 s. For an upward spin configuration in F3GT, the NbSe$_2$ layer functions like a positive-polarity diode and allows non-dissipative transport for the positive cycles only. An opposite rectification mechanism can be achieved by premag-field-driven diode polarity reversal. The V-I curve in Fig.~\ref{fig4}(c) depicts the reversed SDE under a -1 T premag field, with a diode efficiency of 11.1 \%. The intersection of the vertical and horizontal dashed lines denotes the operating point of rectification measurements. 
The reversed rectification characteristic is shown in Fig.~\ref{fig4}(d), with a zero output voltage during negative current cycles. 

We further demonstrate that an XNOR logic gate can be implemented using the highly efficient SDE in the presence of a small active magnetic field. Fig.~\ref{fig5}(a) and (b) present the V-I characteristics of the NbSe$_2$ layer, subjected to applied magnetic fields of -50 mT and 50 mT, respectively. A large non-reciprocal current ($\approx$ 100 $\mu$A) is observed in both cases, shown in the cyan-shaded regions. The diode polarity reverses with the direction of the applied field, which is desirable for logic-gate operation. We choose a time-varying square-wave magnetic field as input A, with a period of 40 s and an amplitude of 50 mT. The 50 mT and -50 mT field levels are defined as ``0" and ``1" states, as illustrated in the top panel of Fig.~\ref{fig5}(a). An electrical bias current of a square waveform, with an amplitude of 230 $\mu$A and a time period of 20 s, is selected as input B of the logic gate. The current amplitude is carefully chosen to be in the non-reciprocal regime of the SDE, as marked by the vertical dashed lines in Fig.~\ref{fig5}(a) and (b). The resulting voltage level of the resistive path is shown in the horizontal dashed lines. The input current waveform and resulting voltage waveform are depicted in the 2nd and 3rd panels of Fig.~\ref{fig5}(c), respectively. The output C of the logic-gate device is denoted by the corresponding resistance, as shown in the last panel of Fig.~\ref{fig5}(c). The high-resistance and zero-resistance outputs are chosen as the ``1" and ``0" state, respectively. At a magnetic field of 50 mT, the reversed-diode characteristic produces a finite-resistance state during positive cycles and a zero-resistive state during negative cycles. Thus, the (1,1) and (1,0) input combinations result in the ``1" and ``0" output states, respectively. This is shown by the black and purple dashed lines in Fig.~\ref{fig5}(c). On the other hand, in a -50 mT magnetic field, a forward-diode behavior allows non-dissipative positive current cycles and resistive negative current cycles. This manifests as a high output state for the (0,0) input state and a low output state for (0,1) input state. This is highlighted by the pink and red dashed lines in Fig.~\ref{fig5}(c), respectively. Collectively, these input-output characteristics describe an XNOR logic operation as summarized in the truth table in the bottom right panel of Fig.~\ref{fig5}(c), demonstrating the applicability of the junction-free NbSe$_2$-based superconducting diode for low-power logic devices. We further perform a long time-series measurement of the rectification and logic operations over more than 100 cycles. Fig.~4 of the supporting information~\cite{SM} shows sustained operations for over 3000 seconds, underlining the robustness of the field-free SDE. 

In conclusion, by using a highly anisotropic magnetic proximity layer, we realize a field-free, junction-free superconducting diode effect in the Ising superconductor NbSe$_2$. The origin of the SDE is associated with finite-momentum Cooper pairs induced by an out-of-plane spin configuration in F3GT layers, which lies parallel to the Ising spins in NbSe$_2$. Thus, the SDE polarity can be switched deterministically by manipulating the spin configuration in F3GT. We illustrate a field-free half-wave diode rectification effect using a square-wave input current bias. The sustained rectification over a long period confirms the stability and durability of the SDE. A zero-field diode efficiency of up to 11.1\% is achieved, increased to 29.7\% in the presence of a 50 mT magnetic field. We further demonstrate XNOR logic gate operation using the magnetic-field- and electric-current-controlled SDE, opening up the prospects of the applicability of junction-free SDE to realize high-efficiency, low-power superconducting logic devices.

{\it Acknowledgements.---} The authors acknowledge IIT Kanpur and the Department of Science and Technology, India,
[Order No. DST/NM/TUE/QM-06/2019 (G)] for financial support. AB thanks PMRF for financial support.

{\it Conflicts of interest.---} The authors declare no conflict of interest.

\bibliography{ref}

@Article{Ma2025,
author={Ma, Jiaxiang
and Wang, Huiyu
and Zhuo, Weizhuang
and Lei, Bin
and Wang, Shuai
and Wang, Wenxiang
and Chen, Xin-Yu
and Wang, Zhen-Yu
and Ge, Binghui
and Wang, Zhen
and Tao, Jing
and Jiang, Kun
and Xiang, Ziji
and Chen, Xian-Hui},
title={Field-free {J}osephson diode effect in {NbSe$_2$} van der {W}aals junction},
journal={Communications Physics},
year={2025},
month={Mar},
day={31},
volume={8},
number={1},
pages={125},
issn={2399-3650},
doi={10.1038/s42005-025-02054-9},
url={https://doi.org/10.1038/s42005-025-02054-9}
}

@article{PhysRevResearch.6.L012046,
  title = {Signature of long-ranged spin triplets across a two-dimensional superconductor/helimagnet van der {W}aals interface},
  author = {Spuri, A. and Nikoli\ifmmode \acute{c}\else \'{c}\fi{}, D. and Chakraborty, S. and Klang, M. and Alpern, H. and Millo, O. and Steinberg, H. and Belzig, W. and Scheer, E. and Di Bernardo, A.},
  journal = {Phys. Rev. Res.},
  volume = {6},
  issue = {1},
  pages = {L012046},
  numpages = {6},
  year = {2024},
  month = {Mar},
  publisher = {American Physical Society},
  doi = {10.1103/PhysRevResearch.6.L012046},
  url = {https://link.aps.org/doi/10.1103/PhysRevResearch.6.L012046}
}

@article{doi:10.1021/acs.nanolett.0c02412,
author = {Indolese, David I. and Karnatak, Paritosh and Kononov, Artem and Delagrange, Raphaëlle and Haller, Roy and Wang, Lujun and Makk, P{\'e}ter and Watanabe, Kenji and Taniguchi, Takashi and Sch{\"o}nenberger, Christian},
title = {Compact {SQUID} Realized in a Double-Layer Graphene Heterostructure},
journal = {Nano Letters},
volume = {20},
number = {10},
pages = {7129-7135},
year = {2020},
doi = {10.1021/acs.nanolett.0c02412},
    note ={PMID: 32872789},
URL = { https://doi.org/10.1021/acs.nanolett.0c02412
}}

@Article{Wu2022,
author={Wu, Heng
and Wang, Yaojia
and Xu, Yuanfeng
and Sivakumar, Pranava K.
and Pasco, Chris
and Filippozzi, Ulderico
and Parkin, Stuart S. P.
and Zeng, Yu-Jia
and McQueen, Tyrel
and Ali, Mazhar N.},
title={The field-free {J}osephson diode in a van der {W}aals heterostructure},
journal={Nature},
year={2022},
month={Apr},
day={01},
volume={604},
number={7907},
pages={653-656},
issn={1476-4687},
doi={10.1038/s41586-022-04504-8},
url={https://doi.org/10.1038/s41586-022-04504-8}
}

@Article{Wu2025,
author={Wu, Si-Li
and Ren, Zhi-Hui
and Yang, Liu
and Wang, Mao-Yuan
and Zhang, Xian-Peng
and Fan, Xiao-Yue
and Zhang, Hao-Chen
and Li, Xiang
and Wang, Gang
and Wang, Chong
and Li, Chuan
and Wang, Zhi-Wei
and Li, Cai-Zhen
and Liao, Zhi-Min
and Yao, Yu-Gui},
title={Efficiency-Tunable Field-Free {J}osephson Diode Effect in {Nb$_3$Cl$_8$} Based van der {W}aals Junctions},
journal={Nano Letters},
year={2025},
month={Dec},
day={24},
publisher={American Chemical Society},
volume={25},
number={51},
pages={17619-17627},
issn={1530-6984},
doi={10.1021/acs.nanolett.5c04336},
url={https://doi.org/10.1021/acs.nanolett.5c04336}
}

@article{doi:10.1021/acs.nanolett.2c01640,
author = {Kang, Kaifei and Berger, Helmuth and Watanabe, Kenji and Taniguchi, Takashi and Forró, László and Shan, Jie and Mak, Kin Fai},
title = {van der {W}aals $\pi$ {J}osephson Junctions},
journal = {Nano Letters},
volume = {22},
number = {13},
pages = {5510-5515},
year = {2022},
doi = {10.1021/acs.nanolett.2c01640},
    note ={PMID: 35736540},
URL = {https://doi.org/10.1021/acs.nanolett.2c01640}
}

@Article{Ai2021,
author={Ai, Linfeng
and Zhang, Enze
and Yang, Jinshan
and Xie, Xiaoyi
and Yang, Yunkun
and Jia, Zehao
and Zhang, Yuda
and Liu, Shanshan
and Li, Zihan
and Leng, Pengliang
and Cao, Xiangyu
and Sun, Xingdan
and Zhang, Tongyao
and Kou, Xufeng
and Han, Zheng
and Xiu, Faxian
and Dong, Shaoming},
title={Van der {W}aals ferromagnetic {J}osephson junctions},
journal={Nature Communications},
year={2021},
month={Nov},
day={12},
volume={12},
number={1},
pages={6580},
doi={10.1038/s41467-021-26946-w},
url={https://doi.org/10.1038/s41467-021-26946-w}
}

@misc{gonzalezsanchez2025,
title={Signatures of edge states in antiferromagnetic van der {W}aals {J}osephson junctions}, 
author={Celia González-Sánchez and Ignacio Sardinero and Jorge Cuadra and Alfredo Spuri and José A. Moreno and Hermann Suderow and Elke Scheer and Pablo Burset and Angelo Di Bernardo and Rubén Seoane Souto and Eduardo J. H. Lee},
year={2025},
eprint={2505.18578},
archivePrefix={arXiv},
primaryClass={cond-mat.supr-con},
url={https://arxiv.org/abs/2505.18578}, 
}

@Article{Jansen2024,
author={Jansen, Thies
and Kochetkova, Ekaterina
and Isaeva, Anna
and Brinkman, Alexander
and Li, Chuan},
title={Josephson coupling across magnetic topological insulator {MnBi$_2$Te$_4$}},
journal={Communications Materials},
year={2024},
month={Oct},
day={08},
volume={5},
number={1},
pages={214},
issn={2662-4443},
doi={10.1038/s43246-024-00649-3},
url={https://doi.org/10.1038/s43246-024-00649-3}
}

@article{
doi:10.1126/sciadv.ads8730,
author = {Enze Zhang  and Zi-Ting Sun  and Zehao Jia  and Jinshan Yang  and Jingyi Yan  and Linfeng Ai  and Ying-Ming Xie  and Yuda Zhang  and Xue-Jian Gao  and Xian Xu  and Shanshan Liu  and Qiang Ma  and Chaowei Hu  and Xufeng Kou  and Jin Zou  and Ni Ni  and Kam Tuen Law  and Shaoming Dong  and Faxian Xiu },
title = {Observation of edge supercurrent in topological antiferromagnet {MnBi$_2$Te$_4$}-based {J}osephson junctions},
journal = {Science Advances},
volume = {11},
number = {20},
pages = {eads8730},
year = {2025},
doi = {10.1126/sciadv.ads8730},
URL = {https://www.science.org/doi/abs/10.1126/sciadv.ads8730}}

@Article{Hu2023,
author={Hu, Guojing
and Wang, Changlong
and Wang, Shasha
and Zhang, Ying
and Feng, Yan
and Wang, Zhi
and Niu, Qian
and Zhang, Zhenyu
and Xiang, Bin},
title={Long-range skin {J}osephson supercurrent across a van der {W}aals ferromagnet},
journal={Nature Communications},
year={2023},
month={Mar},
day={30},
volume={14},
number={1},
pages={1779},
issn={2041-1723},
doi={10.1038/s41467-023-37603-9},
url={https://doi.org/10.1038/s41467-023-37603-9}
}

@article{https://doi.org/10.1002/adma.202513434,
author = {Hu, Guojing and Han, Yechao and Guo, Hui and Lv, Senhao and Gao, Tianqi and Wang, Yunhao and Zhao, Zhen and Zhu, Ke and Qi, Qi and Xian, Guoyu and Zhu, Shiyu and Bao, Lihong and Lin, Xiao and Zhou, Wu and Jiang, Kun and Hu, Jiangping and Yang, Haitao and Gao, Hong-Jun},
title = {Polarity-Reversible Zero-Field Diode Effect in van der {W}aals Ferromagnetic {J}osephson Junction for Logic Operation},
journal = {Advanced Materials},
volume = {38},
number = {9},
pages = {e13434},
doi = {https://doi.org/10.1002/adma.202513434},
url = {https://advanced.onlinelibrary.wiley.com/doi/abs/10.1002/adma.202513434},
year = {2026}
}

@Article{Qiu2023,
author={Qiu, Gang
and Yang, Hung-Yu
and Hu, Lunhui
and Zhang, Huairuo
and Chen, Chih-Yen
and Lyu, Yanfeng
and Eckberg, Christopher
and Deng, Peng
and Krylyuk, Sergiy
and Davydov, Albert V.
and Zhang, Ruixing
and Wang, Kang L.},
title={Emergent ferromagnetism with superconductivity in {Fe(Te,Se)} van der {W}aals {J}osephson junctions},
journal={Nature Communications},
year={2023},
month={Oct},
day={23},
volume={14},
number={1},
pages={6691},
doi={10.1038/s41467-023-42447-4},
url={https://doi.org/10.1038/s41467-023-42447-4}
}

@article{doi:10.1021/acsnano.4c16050,
author = {Hu, Guojing and Wang, Changlong and Lu, Jingdi and Zhu, Yuanmin and Xi, Chuanying and Ma, Xiang and Yang, Yutong and Zhang, Ying and Wang, Shasha and Gu, Meng and Zhang, Jinxing and Lu, Yalin and Cui, Ping and Chen, Guorui and Zhu, Wenguang and Xiang, Bin and Zhang, Zhenyu},
title = {Proximity-Induced Superconductivity in Ferromagnetic {Fe$_3$GeTe$_2$} and {J}osephson Tunneling through a van der {W}aals Heterojunction},
journal = {ACS Nano},
volume = {19},
number = {5},
pages = {5709-5717},
year = {2025},
doi = {10.1021/acsnano.4c16050},
URL = {https://doi.org/10.1021/acsnano.4c16050
}}

@article{doi:10.1021/acsami.3c15363,
author = {Zeng, Xiangyu and Ye, Ge and Yang, Fazhi and Ye, Qikai and Zhang, Liang and Ma, Boyang and Liu, Yulu and Xie, Mengwei and Han, Genquan and Hao, Yue and Luo, Jikui and Lu, Xin and Liu, Yan and Wang, Xiaozhi},
title = {Proximity Effect-Induced Magnetoresistance Enhancement in a {Fe$_3$GeTe$_2$}/{NbSe$_2$}/{Fe$_3$GeTe$_2$} Magnetic Tunnel Junction},
journal = {ACS Applied Materials \& Interfaces},
volume = {15},
number = {49},
pages = {57397-57403},
year = {2023},
doi = {10.1021/acsami.3c15363},
URL = {    
https://doi.org/10.1021/acsami.3c15363
}}

@Article{Nadeem2023,
author={Nadeem, Muhammad
and Fuhrer, Michael S.
and Wang, Xiaolin},
title={The superconducting diode effect},
journal={Nature Reviews Physics},
year={2023},
month={Oct},
day={01},
volume={5},
number={10},
pages={558-577},
issn={2522-5820},
doi={10.1038/s42254-023-00632-w},
url={https://doi.org/10.1038/s42254-023-00632-w}
}

@misc{dibernardo2026,
      title={Van der {W}aals superconducting electronics: materials, devices and circuit integration}, 
      author={Angelo Di Bernardo and Elke Scheer},
      year={2026},
      eprint={2601.07018},
      archivePrefix={arXiv},
      primaryClass={cond-mat.supr-con},
      url={https://arxiv.org/abs/2601.07018}, 
}

@Article{Sarkar2026,
author={Sarkar, Joydip
and Mukherjee, Ayshi
and Basu, Amit
and Kundu, Ritajit
and Kundu, Arijit
and Deshmukh, Mandar M.},
title={van der {W}aals {J}osephson junctions for quantum science and technology},
journal={Nature Reviews Physics},
year={2026},
month={Aug},
day={17},
issn={2522-5820},
doi={10.1038/s42254-026-00970-5},
url={https://doi.org/10.1038/s42254-026-00970-5}
}

@Article{Ando2020,
author={Ando, Fuyuki
and Miyasaka, Yuta
and Li, Tian
and Ishizuka, Jun
and Arakawa, Tomonori
and Shiota, Yoichi
and Moriyama, Takahiro
and Yanase, Youichi
and Ono, Teruo},
title={Observation of superconducting diode effect},
journal={Nature},
year={2020},
month={Aug},
day={01},
volume={584},
number={7821},
pages={373-376},
issn={1476-4687},
doi={10.1038/s41586-020-2590-4},
url={https://doi.org/10.1038/s41586-020-2590-4}
}

@Article{Bauriedl2022,
author={Bauriedl, Lorenz
and B{\"a}uml, Christian
and Fuchs, Lorenz
and Baumgartner, Christian
and Paulik, Nicolas
and Bauer, Jonas M.
and Lin, Kai-Qiang
and Lupton, John M.
and Taniguchi, Takashi
and Watanabe, Kenji
and Strunk, Christoph
and Paradiso, Nicola},
title={Supercurrent diode effect and magnetochiral anisotropy in few-layer {NbSe$_2$}},
journal={Nature Communications},
year={2022},
month={Jul},
day={23},
volume={13},
number={1},
pages={4266},
issn={2041-1723},
doi={10.1038/s41467-022-31954-5},
url={https://doi.org/10.1038/s41467-022-31954-5}
}

@Article{Jeon2022,
author={Jeon, Kun-Rok
and Kim, Jae-Keun
and Yoon, Jiho
and Jeon, Jae-Chun
and Han, Hyeon
and Cottet, Audrey
and Kontos, Takis
and Parkin, Stuart S. P.},
title={Zero-field polarity-reversible {J}osephson supercurrent diodes enabled by a proximity-magnetized {Pt} barrier},
journal={Nature Materials},
year={2022},
month={Sep},
day={01},
volume={21},
number={9},
pages={1008-1013},
issn={1476-4660},
doi={10.1038/s41563-022-01300-7},
url={https://doi.org/10.1038/s41563-022-01300-7}
}

@article{adfm.202311229,
author = {Chen, Pingbo and Wang, Gongqi and Ye, Bicong and Wang, Jinhua and Zhou, Liang and Tang, Zhenzhong and Wang, Le and Wang, Jiannong and Zhang, Wenqing and Mei, Jiawei and Chen, Weiqiang and He, Hongtao},
title = {Edelstein Effect Induced Superconducting Diode Effect in Inversion Symmetry Breaking {MoTe$_2$} {J}osephson Junctions},
journal = {Advanced Functional Materials},
volume = {34},
number = {10},
pages = {2311229},
doi = {https://doi.org/10.1002/adfm.202311229},
url = {https://advanced.onlinelibrary.wiley.com/doi/abs/10.1002/adfm.202311229},
year = {2024}}

@Article{Pal2022,
author={Pal, Banabir
and Chakraborty, Anirban
and Sivakumar, Pranava K.
and Davydova, Margarita
and Gopi, Ajesh K.
and Pandeya, Avanindra K.
and Krieger, Jonas A.
and Zhang, Yang
and Date, Mihir
and Ju, Sailong
and Yuan, Noah
and Schr{\"o}ter, Niels B. M.
and Fu, Liang
and Parkin, Stuart S. P.},
title={Josephson diode effect from {C}ooper pair momentum in a topological semimetal},
journal={Nature Physics},
year={2022},
month={Oct},
day={01},
volume={18},
number={10},
pages={1228-1233},
issn={1745-2481},
doi={10.1038/s41567-022-01699-5},
url={https://doi.org/10.1038/s41567-022-01699-5}
}

@article{Hu2025, 
author = {Guojing Hu and Yechao Han and Weiqi Yu and Senhao Lv and Yuhui Li and Zizhao Gong and Hui Guo and Ke Zhu and Zhen Zhao and Qi Qi and Guoyu Xian and Lihong Bao and Xiao Lin and Jinbo Pan and Shixuan Du and Haitao Yang and Hong-Jun Gao},
title = {Tunable zero-field superconducting diode effect in two-dimensional ferromagnetic/superconducting {Fe$_3$GeTe$_2$/NbSe$_2$} heterostructure},
year = {2025},
journal = {Nano Research},
volume = {18},
number = {1},
pages = {94907068},
url = {https://www.sciopen.com/article/10.26599/NR.2025.94907068},
doi = {10.26599/NR.2025.94907068}
}

@Article{Lyu2021,
author={Lyu, Yang-Yang
and Jiang, Ji
and Wang, Yong-Lei
and Xiao, Zhi-Li
and Dong, Sining
and Chen, Qing-Hu
and Milo{\v{s}}evi{\'{c}}, Milorad V.
and Wang, Huabing
and Divan, Ralu
and Pearson, John E.
and Wu, Peiheng
and Peeters, Francois M.
and Kwok, Wai-Kwong},
title={Superconducting diode effect via conformal-mapped nanoholes},
journal={Nature Communications},
year={2021},
month={May},
day={11},
volume={12},
number={1},
pages={2703},
issn={2041-1723},
doi={10.1038/s41467-021-23077-0},
url={https://doi.org/10.1038/s41467-021-23077-0}
}

@Article{Narita2022,
author={Narita, Hideki
and Ishizuka, Jun
and Kawarazaki, Ryo
and Kan, Daisuke
and Shiota, Yoichi
and Moriyama, Takahiro
and Shimakawa, Yuichi
and Ognev, Alexey V.
and Samardak, Alexander S.
and Yanase, Youichi
and Ono, Teruo},
title={Field-free superconducting diode effect in noncentrosymmetric superconductor/ferromagnet multilayers},
journal={Nature Nanotechnology},
year={2022},
month={Aug},
day={01},
volume={17},
number={8},
pages={823-828},
issn={1748-3395},
doi={10.1038/s41565-022-01159-4},
url={https://doi.org/10.1038/s41565-022-01159-4}
}

@article{PhysRevResearch.5.L022064,
  title = {Magnetic proximity-induced superconducting diode effect and infinite magnetoresistance in a van der {W}aals heterostructure},
  author = {Yun, Jonginn and Son, Suhan and Shin, Jeacheol and Park, Giung and Zhang, Kaixuan and Shin, Young Jae and Park, Je-Geun and Kim, Dohun},
  journal = {Phys. Rev. Res.},
  volume = {5},
  issue = {2},
  pages = {L022064},
  numpages = {7},
  year = {2023},
  month = {Jun},
  publisher = {American Physical Society},
  doi = {10.1103/PhysRevResearch.5.L022064},
  url = {https://link.aps.org/doi/10.1103/PhysRevResearch.5.L022064}
}

@article{doi:10.1021/acsami.5c19869,
author = {Luth, Christopher and Jha, Rajveer and Schalip, Ryan and Sloan, Luke and Disiena, Matthew and Zhang, Hongming and Banerjee, Sanjay K.},
title = {Intrinsic Superconducting Diode Effect Enhancement in {FeSeTe} by Increased Flux Pinning},
journal = {ACS Applied Materials \& Interfaces},
volume = {18},
number = {9},
pages = {14344-14351},
year = {2026},
doi = {10.1021/acsami.5c19869},
URL = { 
https://doi.org/10.1021/acsami.5c19869
}
}

@article{Han2026, 
author = {Yechao Han and Guojing Hu and Zouyouwei Lu and Senhao Lv and Zhen Zhao and Jie Liu and Jinan Shi and Hui Guo and Lihong Bao and Xiaoli Dong and Wu Zhou and Haitao Yang and Xiao Lin and Hong-jun Gao},
title = {Intrinsic field-free superconducting diode effect in a simple van der {W}aals {FeSe} nanosheet},
year = {2026},
journal = {Nano Research},
volume = {19},
number = {7},
pages = {94908621},
url = {https://www.sciopen.com/article/10.26599/NR.2026.94908621},
doi = {10.26599/NR.2026.94908621}
}

@Article{Le2024,
author={Le, Tian
and Pan, Zhiming
and Xu, Zhuokai
and Liu, Jinjin
and Wang, Jialu
and Lou, Zhefeng
and Yang, Xiaohui
and Wang, Zhiwei
and Yao, Yugui
and Wu, Congjun
and Lin, Xiao},
title={Superconducting diode effect and interference patterns in kagome {CsV$_3$Sb$_5$}},
journal={Nature},
year={2024},
month={Jun},
day={01},
volume={630},
number={8015},
pages={64-69},
issn={1476-4687},
doi={10.1038/s41586-024-07431-y},
url={https://doi.org/10.1038/s41586-024-07431-y}
}

@Article{Qi2025,
author={Qi, Shichao
and Ge, Jun
and Ji, Chengcheng
and Ai, Yiwen
and Ma, Gaoxing
and Wang, Ziqiao
and Cui, Zihan
and Liu, Yi
and Wang, Ziqiang
and Wang, Jian},
title={High-temperature field-free superconducting diode effect in {high-T$_\mathrm{c}$} cuprates},
journal={Nature Communications},
year={2025},
month={Jan},
day={09},
volume={16},
number={1},
pages={531},
issn={2041-1723},
doi={10.1038/s41467-025-55880-4},
url={https://doi.org/10.1038/s41467-025-55880-4}
}

@article{PhysRevB.90.081402,
  title = {Origin of the insulating state in exfoliated {high-T$_\mathrm{c}$} two-dimensional atomic crystals},
  author = {Sandilands, L. J. and Reijnders, A. A. and Su, A. H. and Baydina, V. and Xu, Z. and Yang, A. and Gu, G. and Pedersen, T. and Borondics, F. and Burch, K. S.},
  journal = {Phys. Rev. B},
  volume = {90},
  issue = {8},
  pages = {081402(R)},
  numpages = {5},
  year = {2014},
  month = {Aug},
  publisher = {American Physical Society},
  doi = {10.1103/PhysRevB.90.081402},
  url = {https://link.aps.org/doi/10.1103/PhysRevB.90.081402}
}

@article{PhysRevLett.98.057003,
  title = {Penetration Depth Study of Superconducting Gap Structure of {2H-NbSe$_2$}},
  author = {Fletcher, J. D. and Carrington, A. and Diener, P. and Rodi\`ere, P. and Brison, J. P. and Prozorov, R. and Olheiser, T. and Giannetta, R. W.},
  journal = {Phys. Rev. Lett.},
  volume = {98},
  issue = {5},
  pages = {057003},
  numpages = {4},
  year = {2007},
  month = {Feb},
  publisher = {American Physical Society},
  doi = {10.1103/PhysRevLett.98.057003},
  url = {https://link.aps.org/doi/10.1103/PhysRevLett.98.057003}
}

@Article{Xi2016,
author={Xi, Xiaoxiang
and Wang, Zefang
and Zhao, Weiwei
and Park, Ju-Hyun
and Law, Kam Tuen
and Berger, Helmuth
and Forr{\'o}, L{\'a}szl{\'o}
and Shan, Jie
and Mak, Kin Fai},
title={Ising pairing in superconducting {NbSe$_2$} atomic layers},
journal={Nature Physics},
year={2016},
month={Feb},
day={01},
volume={12},
number={2},
pages={139-143},
issn={1745-2481},
doi={10.1038/nphys3538},
url={https://doi.org/10.1038/nphys3538}
}

@Article{Fei2018,
author={Fei, Zaiyao
and Huang, Bevin
and Malinowski, Paul
and Wang, Wenbo
and Song, Tiancheng
and Sanchez, Joshua
and Yao, Wang
and Xiao, Di
and Zhu, Xiaoyang
and May, Andrew F.
and Wu, Weida
and Cobden, David H.
and Chu, Jiun-Haw
and Xu, Xiaodong},
title={Two-dimensional itinerant ferromagnetism in atomically thin {Fe$_3$GeTe$_2$}},
journal={Nature Materials},
year={2018},
month={Sep},
day={01},
volume={17},
number={9},
pages={778-782},
issn={1476-4660},
doi={10.1038/s41563-018-0149-7},
url={https://doi.org/10.1038/s41563-018-0149-7}
}

@misc{SM,
  note = "See Supporting Information, available at www.yyy.com."
}

@article{https://doi.org/10.1002/adfm.75836,
author = {Zapata, Juan C. and Biasi, Emilio De and Sirena, Martin and Kim, Jeehoon and Haberkorn, Nestor},
title = {Field-Free Memory-Programmable Superconducting Diode},
journal = {Advanced Functional Materials},
volume = {36},
number = {49},
pages = {e75836},
doi = {https://doi.org/10.1002/adfm.75836},
url = {https://advanced.onlinelibrary.wiley.com/doi/abs/10.1002/adfm.75836},
year = {2026}
}

@article{https://doi.org/10.1002/adma.202511414,
author = {Li, Jiaxu and Zhang, Zijian and Wang, Shiqi and He, Yu and Lyu, Haochang and Wang, Qiusha and Dong, Bowen and Zhu, Daoqian and Matsuki, Hisakazu and Zhu, Dapeng and Yang, Guang and Zhao, Weisheng},
title = {Field-Free Superconducting Diode Enabled by Geometric Asymmetry and Perpendicular Magnetization},
journal = {Advanced Materials},
volume = {38},
number = {7},
pages = {e11414},
doi = {https://doi.org/10.1002/adma.202511414},
url = {https://advanced.onlinelibrary.wiley.com/doi/abs/10.1002/adma.202511414},
year = {2026}
}

@article{PhysRevLett.130.266003,
  title = {Josephson Diode Effect Induced by Valley Polarization in Twisted Bilayer Graphene},
  author = {Hu, Jin-Xin and Sun, Zi-Ting and Xie, Ying-Ming and Law, K. T.},
  journal = {Phys. Rev. Lett.},
  volume = {130},
  issue = {26},
  pages = {266003},
  numpages = {6},
  year = {2023},
  month = {Jun},
  publisher = {American Physical Society},
  doi = {10.1103/PhysRevLett.130.266003},
  url = {https://link.aps.org/doi/10.1103/PhysRevLett.130.266003}
}

@Article{Diez-Merida2023,
author={D{\'i}ez-M{\'e}rida, J.
and D{\'i}ez-Carl{\'o}n, A.
and Yang, S. Y.
and Xie, Y.-M.
and Gao, X.-J.
and Senior, J.
and Watanabe, K.
and Taniguchi, T.
and Lu, X.
and Higginbotham, A. P.
and Law, K. T.
and Efetov, Dmitri K.},
title={Symmetry-broken {J}osephson junctions and superconducting diodes in magic-angle twisted bilayer graphene},
journal={Nature Communications},
year={2023},
month={Apr},
day={26},
volume={14},
number={1},
pages={2396},
issn={2041-1723},
doi={10.1038/s41467-023-38005-7},
url={https://doi.org/10.1038/s41467-023-38005-7}
}

@article{
doi:10.1126/sciadv.adw6925,
author = {Andrei Kudriashov  and Xiangyu Zhou  and Razmik A. Hovhannisyan  and Alexander S. Frolov  and Leonid Elesin  and Yi Bo Wang  and Ekaterina V. Zharkova  and Takashi Taniguchi  and Kenji Watanabe  and Zheng Liu  and Kostya S. Novoselov  and Lada V. Yashina  and Xin Zhou  and Denis A. Bandurin },
title = {Non-Majorana origin of anomalous current-phase relation and {J}osephson diode effect in Bi$_2$Se$_3$ NbSe$_2$ {J}osephson junctions},
journal = {Science Advances},
volume = {11},
number = {24},
pages = {eadw6925},
year = {2025},
doi = {10.1126/sciadv.adw6925},
URL = {https://www.science.org/doi/abs/10.1126/sciadv.adw6925}}

@Article{Ingla-Aynes2025,
author={Ingla-Ayn{\'e}s, Josep
and Hou, Yasen
and Wang, Sarah
and Chu, En-De
and Mukhanov, Oleg A.
and Wei, Peng
and Moodera, Jagadeesh S.},
title={Efficient superconducting diodes and rectifiers for quantum circuitry},
journal={Nature Electronics},
year={2025},
month={May},
day={01},
volume={8},
number={5},
pages={411-416},
issn={2520-1131},
doi={10.1038/s41928-025-01375-5},
url={https://doi.org/10.1038/s41928-025-01375-5}
}

@Article{Ghosh2024,
author={Ghosh, Sanat
and Patil, Vilas
and Basu, Amit
and {Kuldeep}
and Dutta, Achintya
and Jangade, Digambar A.
and Kulkarni, Ruta
and Thamizhavel, A.
and Steiner, Jacob F.
and von Oppen, Felix
and Deshmukh, Mandar M.},
title={High-temperature {J}osephson diode},
journal={Nature Materials},
year={2024},
month={May},
day={01},
volume={23},
number={5},
pages={612-618},
issn={1476-4660},
doi={10.1038/s41563-024-01804-4},
url={https://doi.org/10.1038/s41563-024-01804-4}
}

@article{
doi:10.1126/science.abl8371,
author = {S. Y. Frank Zhao  and Xiaomeng Cui  and Pavel A. Volkov  and Hyobin Yoo  and Sangmin Lee  and Jules A. Gardener  and Austin J. Akey  and Rebecca Engelke  and Yuval Ronen  and Ruidan Zhong  and Genda Gu  and Stephan Plugge  and Tarun Tummuru  and Miyoung Kim  and Marcel Franz  and Jedediah H. Pixley  and Nicola Poccia  and Philip Kim },
title = {Time-reversal symmetry breaking superconductivity between twisted cuprate superconductors},
journal = {Science},
volume = {382},
number = {6677},
pages = {1422-1427},
year = {2023},
doi = {10.1126/science.abl8371},
URL = {https://www.science.org/doi/abs/10.1126/science.abl8371}}

@article{https://doi.org/10.1002/adfm.202504056,
author = {He, Jiadian and Ding, Yifan and Zeng, Xiaohui and Zhang, Yiwen and Wang, Yanjiang and Dong, Peng and Wu, Yueshen and Cao, Kecheng and Ran, Kejing and Zhou, Xiang and Wang, Jinghui and Chen, Yulin and Watanabe, Kenji and Taniguchi, Takashi and Yu, Shun-Li and Li, Jian-Xin and Wen, Jinsheng and Li, Jun},
title = {Proximity-Induced Superconducting Diode Effect in Antiferromagnetic {M}ott Insulator $\alpha$-{RuCl$_3$}},
journal = {Advanced Functional Materials},
volume = {35},
number = {42},
pages = {2504056},
doi = {https://doi.org/10.1002/adfm.202504056},
url = {https://advanced.onlinelibrary.wiley.com/doi/abs/10.1002/adfm.202504056},
year = {2025}
}

@Article{Lin2022,
author={Lin, Jiang-Xiazi
and Siriviboon, Phum
and Scammell, Harley D.
and Liu, Song
and Rhodes, Daniel
and Watanabe, K.
and Taniguchi, T.
and Hone, James
and Scheurer, Mathias S.
and Li, J. I. A.},
title={Zero-field superconducting diode effect in small-twist-angle trilayer graphene},
journal={Nature Physics},
year={2022},
month={Oct},
day={01},
volume={18},
number={10},
pages={1221-1227},
issn={1745-2481},
doi={10.1038/s41567-022-01700-1},
url={https://doi.org/10.1038/s41567-022-01700-1}
}

@article{10.1021/acsami.6c08036,
    author = {Bera, Alapan and Mukhopadhyay, Soumik},
    title = {Ultra-Low Current-Driven Non-Volatile Magnetization Switching in Inhomogeneous {Fe$_3$GeTe$_2$} via Exchange-Coupled Artificial Domains},
    journal = {ACS Applied Materials \& Interfaces},
    year = {2026},
    month = {08},
    issn = {1944-8244},
    doi = {10.1021/acsami.6c08036},
    url = {https://doi.org/10.1021/acsami.6c08036}
}

@article{PhysRevB.110.224401,
  title = {Anisotropic magnetization dynamics in {Fe$_5$GeTe$_2$} at room temperature},
  author = {Bera, Alapan and Jana, Nirmalya and Agarwal, Amit and Mukhopadhyay, Soumik},
  journal = {Phys. Rev. B},
  volume = {110},
  issue = {22},
  pages = {224401},
  numpages = {15},
  year = {2024},
  month = {Dec},
  publisher = {American Physical Society},
  doi = {10.1103/PhysRevB.110.224401},
  url = {https://link.aps.org/doi/10.1103/PhysRevB.110.224401}
}

@article{wmb3-5r6b,
  title = {Signatures of topological Hall effect in {Fe$_4$GeTe$_2$} nanoflakes},
  author = {Bera, Alapan and Mukhopadhyay, Soumik},
  journal = {Phys. Rev. B},
  volume = {111},
  issue = {22},
  pages = {224420},
  numpages = {8},
  year = {2025},
  month = {Jun},
  publisher = {American Physical Society},
  doi = {10.1103/wmb3-5r6b},
  url = {https://link.aps.org/doi/10.1103/wmb3-5r6b}
}

@article{https://doi.org/10.1002/pssr.70156,
author = {Bera, Alapan and Mukhopadhyay, Soumik},
title = {Study of Magnetoresistance Plateau as a Probe of Spin-Wave Excitations in {Fe$_4$GeTe$_2$}},
journal = {physica status solidi (RRL) – Rapid Research Letters},
volume = {20},
number = {3},
pages = {e70156},
doi = {https://doi.org/10.1002/pssr.70156},
url = {https://onlinelibrary.wiley.com/doi/abs/10.1002/pssr.70156},
eprint = {https://onlinelibrary.wiley.com/doi/pdf/10.1002/pssr.70156},
year = {2026}
}

@Article{Hamill2021,
author={Hamill, Alex
and Heischmidt, Brett
and Sohn, Egon
and Shaffer, Daniel
and Tsai, Kan-Ting
and Zhang, Xi
and Xi, Xiaoxiang
and Suslov, Alexey
and Berger, Helmuth
and Forr{\'o}, L{\'a}szl{\'o}
and Burnell, Fiona J.
and Shan, Jie
and Mak, Kin Fai
and Fernandes, Rafael M.
and Wang, Ke
and Pribiag, Vlad S.},
title={Two-fold symmetric superconductivity in few-layer {NbSe$_2$}},
journal={Nature Physics},
year={2021},
month={Aug},
day={01},
volume={17},
number={8},
pages={949-954},
issn={1745-2481},
doi={10.1038/s41567-021-01219-x},
url={https://doi.org/10.1038/s41567-021-01219-x}
}

@Article{Wang2017,
author={Wang, Hong
and Huang, Xiangwei
and Lin, Junhao
and Cui, Jian
and Chen, Yu
and Zhu, Chao
and Liu, Fucai
and Zeng, Qingsheng
and Zhou, Jiadong
and Yu, Peng
and Wang, Xuewen
and He, Haiyong
and Tsang, Siu Hon
and Gao, Weibo
and Suenaga, Kazu
and Ma, Fengcai
and Yang, Changli
and Lu, Li
and Yu, Ting
and Teo, Edwin Hang Tong
and Liu, Guangtong
and Liu, Zheng},
title={High-quality monolayer superconductor {NbSe$_2$} grown by chemical vapour deposition},
journal={Nature Communications},
year={2017},
month={Aug},
day={30},
volume={8},
number={1},
pages={394},
issn={2041-1723},
doi={10.1038/s41467-017-00427-5},
url={https://doi.org/10.1038/s41467-017-00427-5}
}

@article{PhysRevResearch.4.013188,
  title = {Orbital-selective two-dimensional superconductivity in {2H-NbSe$_2$}},
  author = {Bi, Xiangyu and Li, Zeya and Huang, Junwei and Qin, Feng and Zhang, Caorong and Xu, Zian and Zhou, Ling and Tang, Ming and Qiu, Caiyu and Tang, Peizhe and Ideue, Toshiya and Nojima, Tsutomu and Iwasa, Yoshihiro and Yuan, Hongtao},
  journal = {Phys. Rev. Res.},
  volume = {4},
  issue = {1},
  pages = {013188},
  numpages = {8},
  year = {2022},
  month = {Mar},
  publisher = {American Physical Society},
  doi = {10.1103/PhysRevResearch.4.013188},
  url = {https://link.aps.org/doi/10.1103/PhysRevResearch.4.013188}
}

@Article{Cho2021,
author={Cho, Chang-woo
and Lyu, Jian
and Ng, Cheuk Yin
and He, James Jun
and Lo, Kwan To
and Chareev, Dmitriy
and Abdel-Baset, Tarob A.
and Abdel-Hafiez, Mahmoud
and Lortz, Rolf},
title={Evidence for the {F}ulde-{F}errell-{L}arkin-{O}vchinnikov state in bulk {NbS$_2$}},
journal={Nature Communications},
year={2021},
month={Jun},
day={16},
volume={12},
number={1},
pages={3676},
issn={2041-1723},
doi={10.1038/s41467-021-23976-2},
url={https://doi.org/10.1038/s41467-021-23976-2}
}

@Article{Ge2015,
author={Ge, Jian-Feng
and Liu, Zhi-Long
and Liu, Canhua
and Gao, Chun-Lei
and Qian, Dong
and Xue, Qi-Kun
and Liu, Ying
and Jia, Jin-Feng},
title={Superconductivity above 100 {K} in single-layer {FeSe} films on doped {SrTiO$_3$}},
journal={Nature Materials},
year={2015},
month={Mar},
day={01},
volume={14},
number={3},
pages={285-289},
issn={1476-4660},
doi={10.1038/nmat4153},
url={https://doi.org/10.1038/nmat4153}
}

@article{PhysRevLett.130.046702,
  title = {Revealing the Origin of Time-Reversal Symmetry Breaking in {Fe}-Chalcogenide Superconductor {FeTe$_\mathrm{1-x}$Se$_\mathrm{x}$}},
  author = {Farhang, Camron and Zaki, Nader and Wang, Jingyuan and Gu, Genda and Johnson, Peter D. and Xia, Jing},
  journal = {Phys. Rev. Lett.},
  volume = {130},
  issue = {4},
  pages = {046702},
  numpages = {6},
  year = {2023},
  month = {Jan},
  publisher = {American Physical Society},
  doi = {10.1103/PhysRevLett.130.046702},
  url = {https://link.aps.org/doi/10.1103/PhysRevLett.130.046702}
}

@Article{sym12091402,
AUTHOR = {Kreisel, Andreas and Hirschfeld, Peter J. and Andersen, Brian M.},
TITLE = {On the Remarkable Superconductivity of {FeSe} and Its Close Cousins},
JOURNAL = {Symmetry},
VOLUME = {12},
YEAR = {2020},
NUMBER = {9},
ARTICLE-NUMBER = {1402},
URL = {https://www.mdpi.com/2073-8994/12/9/1402},
ISSN = {2073-8994},
DOI = {10.3390/sym12091402}
}

@Article{Tan2018,
author={Tan, Cheng
and Lee, Jinhwan
and Jung, Soon-Gil
and Park, Tuson
and Albarakati, Sultan
and Partridge, James
and Field, Matthew R.
and McCulloch, Dougal G.
and Wang, Lan
and Lee, Changgu},
title={Hard magnetic properties in nanoflake van der {W}aals {Fe$_3$GeTe$_2$}},
journal={Nature Communications},
year={2018},
month={Apr},
day={19},
volume={9},
number={1},
pages={1554},
issn={2041-1723},
doi={10.1038/s41467-018-04018-w},
url={https://doi.org/10.1038/s41467-018-04018-w}
}

\end{document}